\documentclass[12pt]{iopart}
\usepackage{graphicx}% Include figure files
\newcommand\avg[1]{\langle #1 \rangle}
\usepackage{iopams}  
\newcommand{\lam}{\lambda}

\begin{document}
\title[Optimal and Pessimal scale-free topologies] {Network
  synchronization: Optimal and Pessimal scale-free topologies}
\author{Luca Donetti} \address{Departamento de Electr\'onica y Tecnolog{\'\i}a 
de Computadores and \\
  Instituto de F{\'\i}sica Te{\'o}rica y Computacional Carlos I,
  Facultad de Ciencias, Universidad de Granada, 18071 Granada, Spain}
\author{Pablo I. Hurtado and Miguel A. Mu\~noz} \address{ Departamento
  de Electromagnetismo y F{\'\i}sica de la
  Materia and Instituto Carlos I de F{\'\i}sica Te{\'o}rica y Computacional \\
  Facultad de Ciencias, Universidad de Granada, 18071 Granada, Spain}
\ead{mamunoz@onsager.ugr.es}
\begin{abstract}

  By employing a recently introduced optimization algorithm we
  explicitely design optimally synchronizable (unweighted) networks
  for any given scale-free degree distribution. We explore how the
  optimization process affects degree-degree correlations and observe
  a generic tendency towards disassortativity. Still, we show that
  there is not a one-to-one correspondence between synchronizability
  and disassortativity. On the other hand, we study the nature of
  optimally un-synchronizable networks, that is, networks whose
  topology minimizes the range of stability of the synchronous state.
  The resulting ``pessimal networks'' turn out to have a highly
  assortative string-like structure. We also derive a rigorous lower
  bound for the Laplacian eigenvalue ratio controlling
  synchronizability, which helps understanding the impact of degree
  correlations on network synchronizability.
 
\end{abstract}
%Uncomment for PACS numbers title message
\vspace{2pc}
\noindent{\it Keywords}: Article preparation, IOP journals
% Uncomment for Submitted to journal title message
\submitto{\JPA}
% Comment out if separate title page not required
\maketitle

\section{Introduction}

Synchronizability is one of the currently leading problems in the
fast-growing field of Complex Networks \cite{Reviews}. A number of
studies have been devoted to scrutinize which network topologies are
more prone to sustain a stable globally-synchronized state of generic
oscillators defined at each of its nodes
\cite{Pecora,Nish,Hong,Motter,Bocca,Atay,Entangled,Ramanujan}.  This
question is of broad interest since many complex systems in fields
ranging from physics, biology, computer science, or physiology, can be
seen as networks of coupled oscillators, whose functionality depends
crucially on the network ability to maintain a synchronous oscillation
pattern.  In addition, it has been shown that networks with good
synchronizability are also ``good'' for (i) fast random walk spreading
and therefore for efficient communication \cite{Entangled}, (ii)
searchability in the presence of congestion \cite{Catalans}, (iii)
robustness in the absence of privileged hubs \cite{Maritan}, (iv)
performance of neural networks \cite{Neural}, (v) generating consensus
in social networks, etc.  Another related and important problem that
has received a lot of attention, but that we will not study here, is
the dynamics {\it towards} synchronized states (see for example
\cite{dynamics}).

In general terms, we can say that the degree of synchronizability is
high when all the different nodes in a given network can ``talk
easily'' to each other, or information packets can travel efficiently
from any starting node to any target one.  It was first observed that
adding some extra links to an otherwise regular lattice in such a way
that a small-world topology \cite{Reviews} is generated, enhances
synchronizability \cite{Pecora}. This was attributed to the fact that
the node-to-node average distance diminishes as extra links are added.
Afterwards, heterogeneity in the degree distribution was shown to
hinder synchronization in networks of symmetrically coupled
oscillators, leading to the so called ``paradox of heterogeneity'' as
heterogeneity is known to reduce in average the node-to-node distance
but still it suppresses synchronizability. The effect of other
topological features as betweenness centrality, correlation in the
degree distribution and clustering has been also analyzed.  For
example, it has been shown that the presence of weighted links (rather
than uniform ones) and asymmetric couplings do enhance further the
degree of synchronizability \cite{Motter,Bocca}, but here we focus on
un-weighted and un-directed links.

Certainly, the main breakthrough was made by Barahona and Pecora
\cite{Pecora} who, in a series of papers, established a criterion
based on spectral theory to determine the stability of synchronized
states under very general conditions. Their main contribution is to
link graph spectral properties with network dynamical properties.  In
particular, they considered the Laplacian matrix, encoding the network
topology, and showed that the degree of synchronizability (understood
as the range of stability of the synchronous state) is controlled by
the ratio between its largest eigenvalue ($\lambda_N$) and the
smallest non-trivial one ($\lambda_2$), i.e. $Q=\lambda_N/\lambda_2$,
where $N$ is the total number of nodes \cite{eigen}. The smaller $Q$
the better the synchronizability.

Note that, as the range of variability of $\lam_N$ is quite limited
(it is directly related to the maximum connectivity
\cite{Mohar,Graphs}), minimizing $Q$ is almost
equivalent to maximizing the denominator $\lam_2$ (i.e the {\it
  spectral gap}) when the degree distribution is kept fixed.

It is worth noticing that, even if the eigenratio $Q$ can be related
to (or bounded by) topological properties such as the ones cited above
(average path length, betweenness centrality, etc.), none of these
provides with a full characterization of a given network and therefore
they are not useful to determine {\it strict} criteria for
synchronizability \cite{Atay}.  Nevertheless, they can be very helpful
as long as they give easy criteria to determine in a {\it rough} way
synchronizability properties, without having to resort to lengthly
eigenvalue calculations.

In a couple of recent papers, we tackled the problem of finding the
optimally synchronizable topology, given a fixed number of nodes and
edges linking them, and assuming symmetric and un-weighted links
\cite{Entangled,Ramanujan}.  The strategy we followed was to implement
a simulated annealing algorithm \cite{SA} with a cost-function given
by $Q$; starting with a random topology with $N$ nodes and $L$ links,
random rewirings that decrease the value of $Q$ are accepted with
larger probability than those increasing $Q$ (for more details see
\cite{Entangled,Ramanujan}), until eventually a stationary (optimal or
close to optimal) network is generated. Employing this optimization
algorithm, we identified the family of ``optimal network topologies''
which we called {\it entangled networks}.

The main topological trait of entangled networks is the absence of
bottlenecks and hubs; all sites are very much alike and the links form
very intricate structures, which lead typically to (i) the absence of
a well-defined community structure, (ii) poor modularity, and (iii)
large shortest-loops.  In this way, every single site is close to any
other one owing to the existence of a very ``democratic'' or entangled
structure in which properties such as site-to-site distance,
betweenness, and minimum-loop-size are very homogeneously distributed
(see \cite{Entangled,Ramanujan}).
%One can say that entangled networks
%are ``super-homogeneous''.

Entangled networks were identified as {\it Ramanujan graphs} and they
have been related to other interesting concepts in graph theory as
{\it expanders} \cite{Sarnak} and cage-graphs (see \cite{Ramanujan}
and references therein).  These are used profusely in computer science
and are under current intense study in the mathematical literature.
For example, expanders and Ramanujan graphs are very useful in the
design of efficient communication networks, construction of
error-correcting codes, or de-randomization of random algorithms
\cite{Sarnak,Ramanujan}. These applications greatly amplify the
relevance of entangled networks in different contexts.

Despite of their mathematical beauty and excellent performance in
network-design, entangled topologies are not easily found in
biological, social, or any other ``real-life'' networks. An exception
are some food-webs, for which topologies very similar to entangled
ones have been reported \cite{Estrada}. As argued in
\cite{Entangled,Ramanujan} the rarity in Nature of entangled networks
comes from the fact that they emerge out of a {\it global}
optimization process not easily fulfilled by means of any dynamical
simple mechanism in growing networks where, usually, only {\it local}
information is available.

Instead, real complex networks in very different contexts have been
shown to exhibit, rather generically, scale-free degree distributions.
These are much more heterogeneous than entangled topologies.  Keeping
this in mind, in this paper we explore the question of (global)
optimization of synchronizability within the realm of scale-free
networks with a fixed degree-distribution.

In particular, constraining our optimization algorithm to preserve a
scale-free architecture, we are able to find the optimally
synchronizable networks and study the emergence of non-trivial
degree-degree correlations. This study is related to previous works by
Sorrentino, di Bernardo and others \cite{Sorrentino}, who argued that
disassortative networks (in which nodes with similar degrees tend to
be {\it not} connected among themselves) \cite{asdis} are more
synchronizable that assortative ones (where nodes with similar degrees
tend to be connected).

Our study differs from previous ones in that (i) we derive a rigorous
lower-bound for $Q$ in terms of a parameter measuring degree-degree
correlations and (ii) we explicitly design optimal networks with a
given degree-distribution and, by doing so, we verify that even if it
is true that more disassortative networks typically exhibit better
synchronizability, this is {\it not} always the case.

Finally, we also face the question of which are the {\it pessimal}
networks for synchronization purposes. Actually, in some applications,
synchronization (or consensus, or complete homogenization) are not
desirable properties. This might be the case, for example, in neural
networks for which global synchronization implies epileptic-like
activity \cite{Glass}. The question of how topology can hinder such
states is both pertinent and relevant and also, it can give further
insight on the key structural features of synchronization.  With this
goal in mind, we revert the optimization algorithm, and define an
inverse optimization process just by maximizing $Q$ (rather than
minimizing it), We analyze the topology of the resulting pessimal (or
optimally un-synchronizable) networks.

 \section{A rigorous upper bound for the spectral gap}

 In this section we derive an upper bound for the spectral graph in
 terms of the correlation coefficient $r$. This coefficient was
 introduced by Newman in \cite{asdis} to quantify the tendency of
 nodes with similar degrees, $k$, to be connected between themselves.
 In particular, calling $N$ the number of nodes and $L$ the total
 number of links ($L=N \avg{k} /2$) the correlation coefficient $r$
 can be computed as (see \cite{asdis} for more details):
\begin{equation}
  \label{r}
  r = \frac{ L^{-1}\sum_{i \sim j}k_ik_j - [L^{-1}\sum_{i \sim
      j}\frac12(k_i+k_j)]^2 } {L^{-1}\sum_{i \sim j}\frac12(k_i^2+k_j^2) 
    - [L^{-1}\sum_{i \sim j}\frac12(k_i+k_j)]^2 } \quad.
\end{equation}
where $\sum_{i \sim j}$ stands for the sum over links (i.e. over all
nodes i and j connected by a link; every link is counted only once).
This parameter takes positive (negative) values for assortative
(disassortative) configurations.

Defining the Laplacian matrix as ${\cal L}_{ij}=-1 \ (0)$ if nodes $i$
and $j$ are connected (disconnected), and ${\cal L}_{ii}=k_i$, we can
obtain an upper bound for $Q$ by recalling that the first non-trivial
Laplacian eigenvalue $\lambda_2$ can be expressed as \cite{Mohar}:
\begin{equation}
  \label{eq:l2}
  \lambda_2 = 2 N \min_{{\bf f} \in \Phi} \frac{\sum_{i \sim j} (f_i
    -f_j)^2 } {\sum_i \sum_j (f_i - f_j)^2} \quad ,
\end{equation}
where $\Phi$ is is the set of all possible non-constant vectors (in
the space in which the Laplacian operator acts). Taking ${\bf
  f}=\{k_i, \ i=1\ldots N\}$, which is one possible vector out of the
set $\Phi$, we obtain:
\begin{equation}
  \label{eq:l2b}
  \lambda_2 \leq  2 N  \frac{\sum_{i \sim j} (k_i - k_j)^2 } {\sum_i
    \sum_j (k_i - k_j)^2} = \frac{A}{\avg{k^2} - \avg{k}^2} \quad, 
\end{equation}
where $A$ is defined as: $ A = \frac1N \sum_{i \sim j} (k_i - k_j)^2.$
A different selection of the vector would lead to a different
inequality.  The advantage of our choice is that the obtained bound
can be related to the correlation coefficient $r$, even if it is not
guaranteed that it provides a tight bound.

The different terms in the numerator and denominator of
equation~(\ref{r}) can be written as
\begin{eqnarray}
&& \frac1L \sum_{i \sim j}\frac12(k_i+k_j) = \frac1{N \avg{k}} \sum_i k_i^2 =
  \frac{\avg{k^2}}{\avg{k}}, \nonumber \\ && \frac1L \sum_{i \sim
  j}\frac12(k_i^2+k_j^2) = \frac1{N \avg{k}} \sum_i k_i^3 =
  \frac{\avg{k^3}}{\avg{k}}, \nonumber \\ && \frac1L \sum_{i \sim j} k_ik_j =
  \frac1{L} \sum_{i \sim j} \left( \frac12 (k_i^2+k_j^2) - \frac12 (k_i -
  k_j)^2 \right) = \frac{\avg{k^3} - A}{\avg{k}} \quad.
\end{eqnarray}
By substituting these expressions in Eq.~(\ref{r}) and rearranging
them we readily obtain:
\begin{equation}
  A = (1-r) \frac{\avg{k}\avg{k^3} - \avg{k^2}\avg{k^2}}{\avg{k}} \quad,
\end{equation}
and finally
\begin{equation}
  \lambda_2 \leq (1-r) \frac{\avg{k}\avg{k^3} - \avg{k^2}\avg{k^2}}
  {\avg{k} (\avg{k^2} - \avg{k}^2)}  \quad,
\label{ineq}
\end{equation}
which provides a rigorous upper bound for the spectral gap in terms of
$r$.  Observe that the more negative the value of $r$ (i.e. the more
disassortative the network) the larger the upper bound for $\lambda_2$
and, therefore, $Q$ is allowed to take smaller values, and the
corresponding network can be more synchronizable.
 
Summing up, the inequality Eq.(\ref{ineq}) establishes that, as a rule
of thumb, disassortative networks are more prone to have stable
synchronized states than assortative ones, in agreement with previous
results \cite{Sorrentino}.  As a word of caution, let us underline
that this does {\it not} imply that, given a fixed degree
distribution, any disassortative network is better synchronizable than
any assortative one, as we will illustrate in the following section.

A similar result to ours has been recently derived \cite{Sorrentino}.
In particular, upper and lower bounds for $\lambda_2$ were obtained in
terms of a parameter $\hat{r}$ quantifying the degree-degree
correlation ($\hat{r}$ is a simplified version of the more detailed
one, $r$, defined by Eq.(\ref{r})). These upper and lower bounds were
derived elaborating upon known bounds for the spectral gap in terms of
the Cheeger constant \cite{Graphs,Mohar}.  In order to obtain them,
the authors implicitly assume that, for a fixed degree distribution,
the Cheeger constant is an uni-parametric function of $\hat{r}$.
However, given a fixed $\hat{r}$, as this parameter does not specify
completely the graph topology \cite{Atay}, different graphs with
different Cheeger constants can be constructed.  Therefore, the
derivation of the bounds in \cite{Sorrentino} involves some type of
mean-field-like approximation, while the upper bound here has been
obtained in a rigorous way.

\section{Optimal and pessimal network design}

In this section we describe the optimization algorithm suitable for
finding the network topology which extremizes the stability range of a
global synchronous state in networks subject to a topological
constraint: a fixed degree distribution. In particular, we apply this
method to networks with scale-free topology and analyze the
degree-degree correlations of the resulting (extremized) graphs.

The algorithm is a modified simulated-annealing \cite{Entangled} aimed
at minimizing a cost function ${\cal F}(Q)$, where
$Q=\lambda_N/\lambda_2$. A detailed description of the algorithm can
be found either in \cite{Entangled} or in \cite{Ramanujan}.  It yields
networks for which the synchronizability is close to extremal (i.e.
maximum or minimum), depending on the selected cost function ${\cal
  F}(Q)$.  In particular, setting ${\cal F}(Q)=Q$ one gets networks
with {\it optimal} synchronizability while choosing ${\cal F}(Q)=-Q$
the optimization procedure yields what we call {\it pessimal}
networks.

\begin{figure}
  \centering{\includegraphics[width=.35 \textwidth]{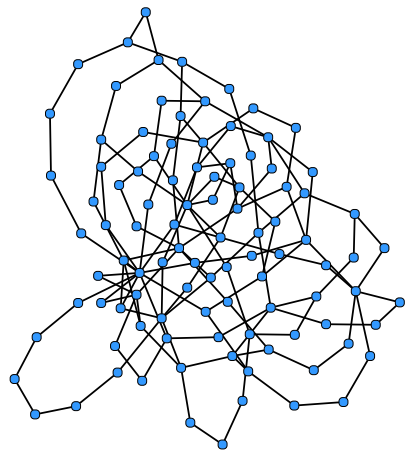}
    \caption{\label{red} Uncorrelated scale-free network (with
      $\gamma=3$, $N=100$, and minimum connectivity $k=2$) from which
      the extremization processes leading to the networks in Fig.
      \ref{opt-pes} is started.}}
\end{figure}

\begin{figure}
    \centering{
    \includegraphics[height=6.0cm]{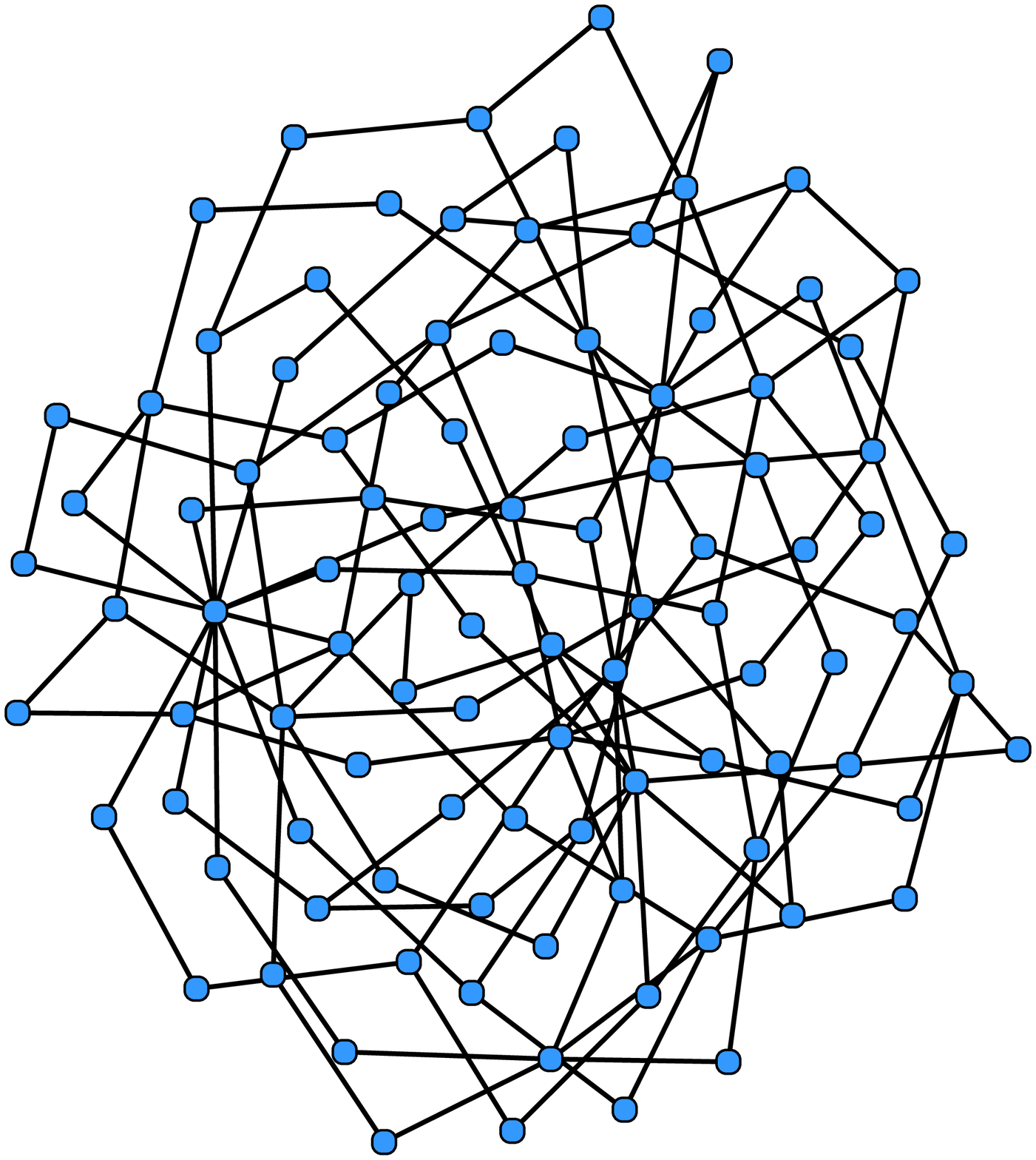}
    \includegraphics[height=6.0cm]{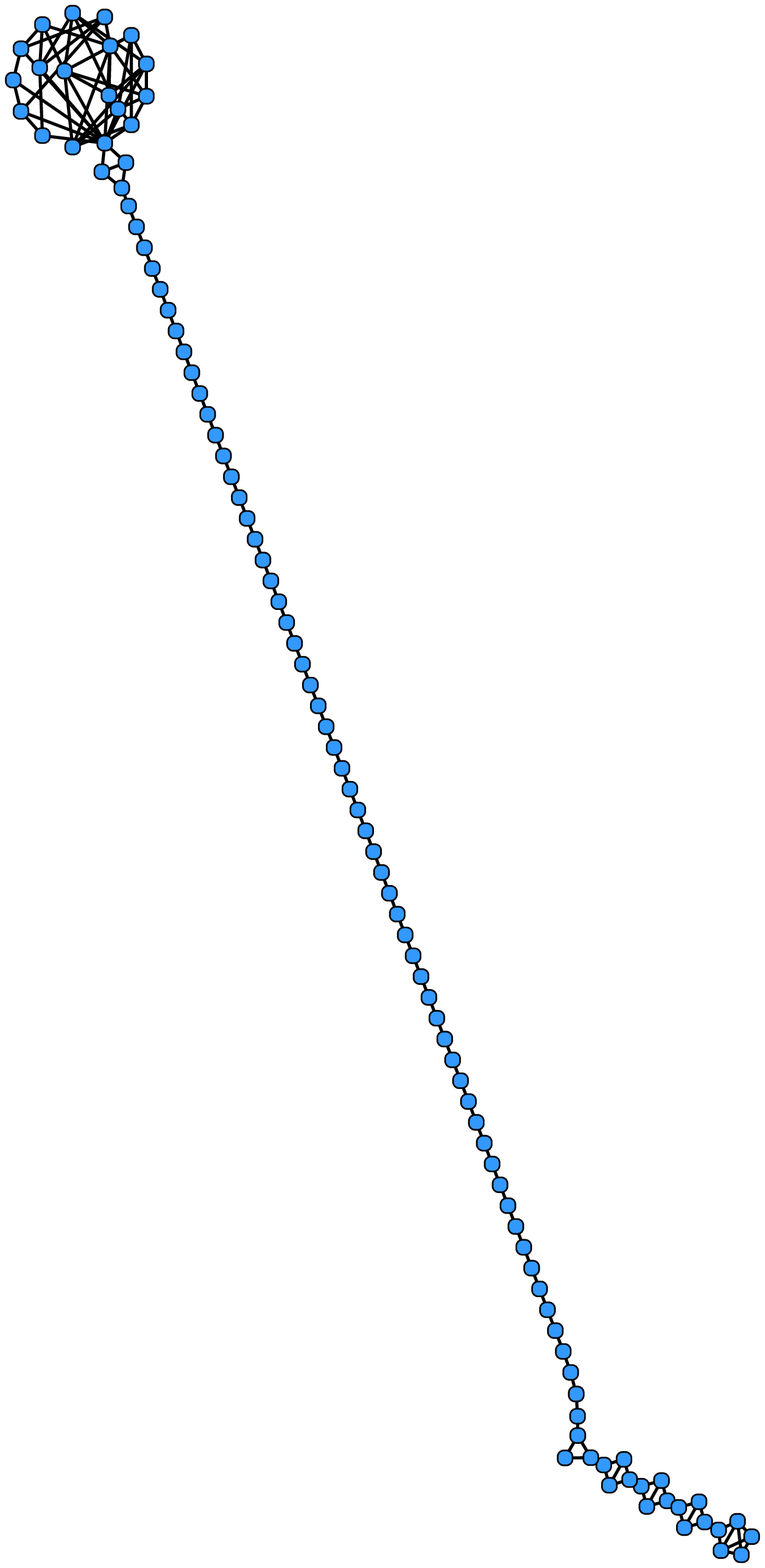}
    \caption{\label{opt-pes} Left: Optimal scale-free network obtained
      from that in Fig.  \ref{red} via minimization of $Q$. Right:
      Pessimal scale-free network obtained from that in Fig. \ref{red}
      via maximization of $Q$. The degree distributions of these two
      networks and the one in Fig. \ref{red} are identical.}
  }
\end{figure}

The simulated-annealing rewiring process starts from networks
generated using the configuration model \cite{conf}; in particular, it
starts from connected networks with $N$ nodes and $L$ links, such that
their degree distribution sample a power law $P(k)\sim k^{-\gamma}$,
$\gamma > 0$, with trivial (random) degree-degree correlation between
neighboring nodes (see Fig.\ref{red}). All the results in what follows
correspond to $\gamma=3$.
The graphs emerging out of the $Q$ and $-Q$ minimization processes,
starting from the network in Fig.~\ref{red}, are depicted in
Fig.\ref{opt-pes}.
Naked eye inspection reveals the enormous differences between optimal
and pessimal topologies. While the optimal ones resemble very much the
very intricate and as-homogeneous-as-possible topology of entangled
networks, pessimal topologies are as chain-like as possible, with two
non-linear ``heads'' at both extremes, necessary to preserve the
scale-free topology constraint. Note that large values of $Q$ imply
small values of $\lambda_2$ and therefore, following the criterion for
graph (bi)partitioning described for instance in \cite{GN}, pessimal
graphs have to be easily divisible into two parts by cutting an
as-small-as-possible number of links. This is, indeed, achieved in an
optimal way for linear (chain-like) topologies.

Let us remark, that different initial conditions with the same
scale-free distribution, lead to outputs indistinguishable
statistically from the ones in Fig.\ref{opt-pes}, rendering robust the
previous results.

To put these observations under a more quantitative basis, we measure
degree correlations using (i) the average degree ${\bar k}_{nn}(k)$ of
the neighbors of a node with degree $k$,
% , defined as
% \begin{equation} {\bar k}_{nn}(k)=\sum_{i\sim j}
%   \left[k_i\delta(k_j-k) + k_j\delta(k_i-k)\right] \ ,
% \label{kmed}
% \end{equation}
% where $\sum_{i\sim j}$ was defined in the previous section, 
and (ii) the correlation coefficient $r$ given by Eq.(\ref{r}).
\begin{figure}
\begin{center}
\includegraphics[width=.55 \textwidth]{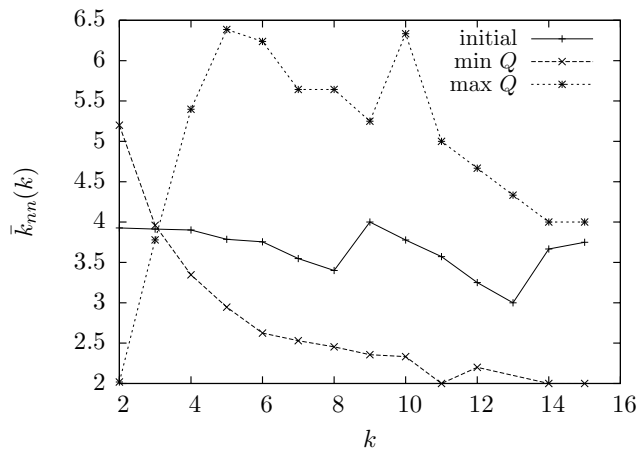}
\caption{Ensemble average of $\bar{k}_{nn}(k)$ defined as
  the mean degree of the neighbors of a node with degree $k$, for (i)
  the initial uncorrelated scale-free networks with $\gamma=3$, (ii)
  the optimal scale-free networks resulting from the simulated
  annealing algorithm, and (iii) the pessimal scale-free networks
  obtained after maximization of $Q$. Results are averaged over an
  ensemble of $10$ different networks of sizes $N=100$.}
\label{pkk}
\end{center}
\end{figure}
Fig.\ref{pkk} shows ${\bar k}_{nn}(k)$, averaged over $10$ different
realizations, for initially uncorrelated networks (see, as an example,
Fig.  \ref{red}) as well as for the final optimal (Fig.
\ref{opt-pes}, left) and pessimal (Fig. \ref{opt-pes}, right)
networks.  It reveals that optimally-synchronizable scale-free
networks tend to display disassortative mixing (high-degree nodes tend
to be connected with low-degree ones) while, on the contrary, pessimal
scale-free networks tend to be assortative. This result agrees with
the tendency predicted by the bound on the spectral gap in Eq.
(\ref{ineq}) as well as with previous results \cite{Sorrentino}.
Actually, one could have anticipated these conclusions knowing that a
network with good synchronization properties is also able to
efficiently communicate any two nodes \cite{Entangled,Ramanujan}. In
this sense, disassortative mixing, in which low connected nodes are
preferentially linked to hubs which act as information distributors,
seems most efficient.

On the other hand, pessimally synchronizable networks resulting from
the minimization of $-Q$ (or, equivalently, maximization of $Q$), tend
to exhibit assortative mixing, i.e.  ${\bar k}_{nn}(k)$ grows with
$k$, at least up to a finite-size cutoff $k^*$, as shown in Fig.
\ref{pkk}.  The origin of such a cutoff is evident after realizing
that the probability of having large hubs connected to other very
large hubs must go to zero since the total number of links present in
the system is finite.  Obviously, the cutoff grows with system size
and diverges asymptotically.  The highly assortative chain-like
topology of pessimal networks can be understood by the necessity of
hampering the efficient communication between any two nodes in the
system. This is achieved by maximizing the distance between any two
hubs by interposing between them a linear chain of poorly-connected
nodes.
\begin{figure}\centering{
\includegraphics[height=8.5cm]{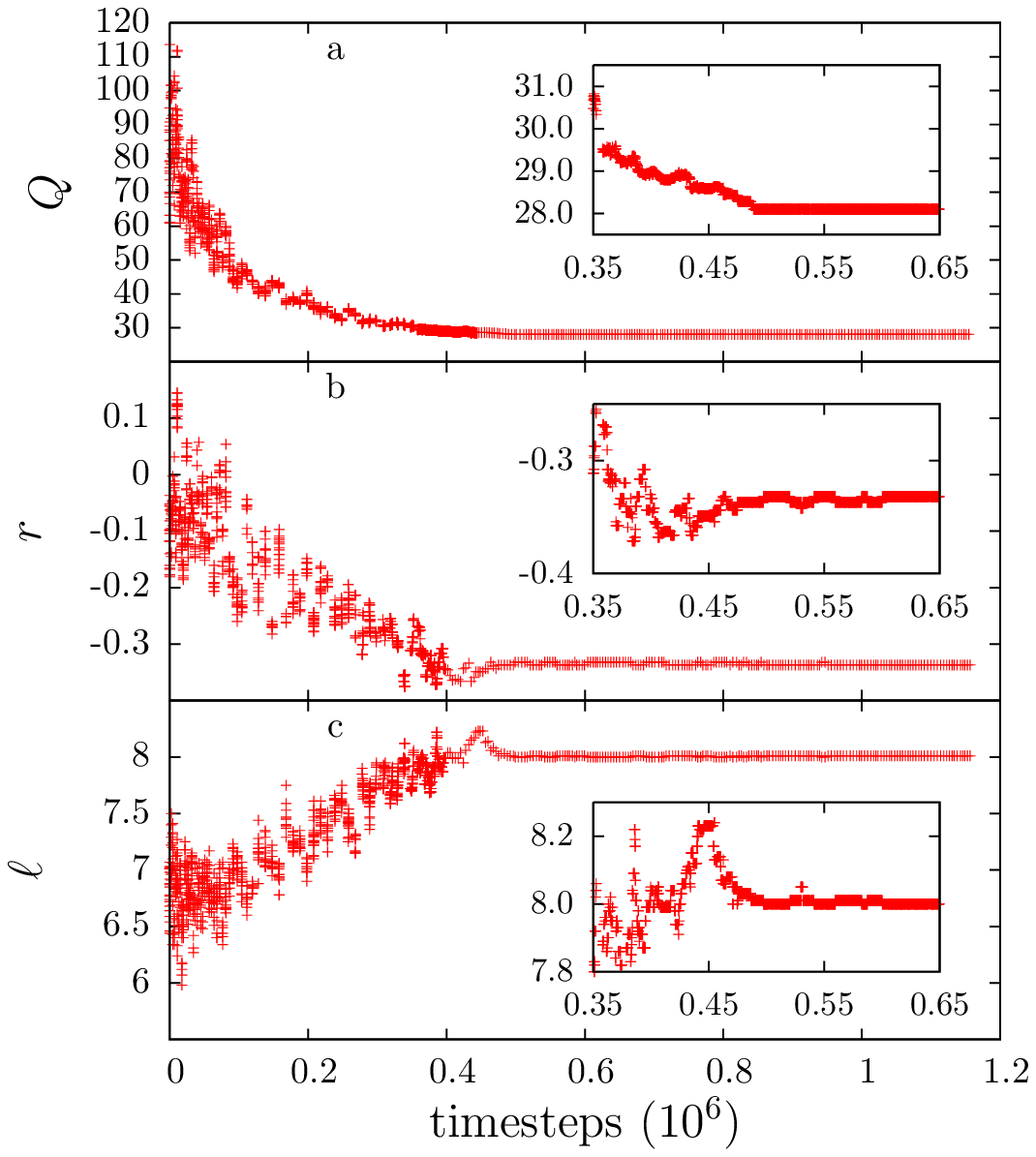}
\includegraphics[height=8.5cm]{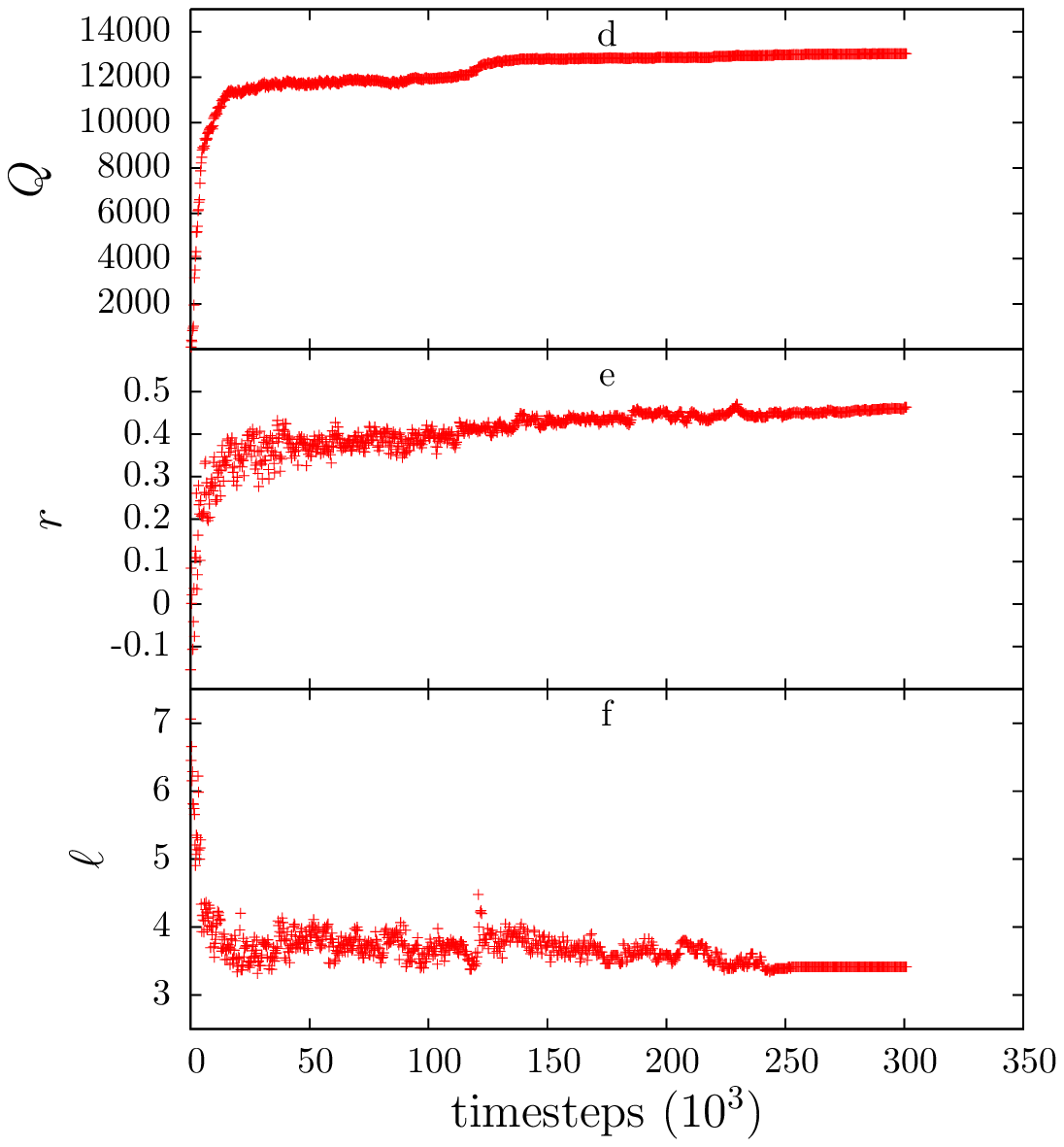}
\caption{\label{evol} Time evolution of different quantities during
  the minimization (left) and maximization (right) procedure of a
  scale-free network with $\gamma=3$, $N=100$ nodes. (a) and (d)
  Eigenvalue ratio $Q$, (b) and (e) correlation coefficient $r$, (c)
  and (f) shortest-loop average length $\ell$. The insets show a zoom
  of the corresponding curves. Note that $r$ (resp. $\ell$) exhibits a
  minimum (resp. maximum) during the algorithmic evolution which does
  not correspond to its optimal value.
% This shows that, though
%  disassortativity and large shortest-loops favor in general
%  synchronizability, their extremization does not guarantee optimal
%  stability of the synchronous state.
}}
\end{figure}

Let us underline that the above observation on the effect of
disassortative (assortative) mixing does not necessarily imply that
maximizing disassortativity (assortativity) leads to optimal
(pessimal) synchronizability. This is illustrated in Figs.
\ref{evol}.a-b (Figs. \ref{evol}.d-e), which plot respectively the
time evolution of the eigenratio $Q$ and the correlation coefficient
$r$ during the optimization processes. The figure regarding
optimization shows that the eigenratio $Q$ is not a monotonic function
of $r$. This fact is made explicit in the insets to Figs.
\ref{evol}.a-b: the asymptotic minimun value of $Q$ does not
correspond to the network for which $r$ is minimum (obtained in the
example shown around $t=4\times 10^5$ steps).  This points out that,
despite being a good indicator of synchronizability, disassortativity
cannot be regarded as an unique topological measure of the stability
of the synchronous state \cite{Atay}.  Moreover, further correlations
apart from the observed assortativity/disassortativity are built up
during the optimization process. In Fig. \ref{evol}.c we plot the time
evolution of the shortest-loop average length $\ell$, defined as the
average over all nodes of the shortest loop passing through each node;
it shows that $\ell$ grows during the minimization of $Q$, but again
it exhibits a maximum before reaching its optimal-topology value. The
tendency towards forming large loops for optimally synchronizable
networks was reported before for entangled networks, where the
smallest loops tend to be as large as possible
\cite{Entangled,Ramanujan}.

At this point we want to emphasize that the approach we have
undertaken here is a constructive one, as opposed to that in
\cite{Sorrentino}, where different random networks with predefined
degree distribution and correlations are explored to analyze how the
externally-imposed assortative or disassortative correlation affects
the eigenratio $Q$ of the resulting networks. Results here complement
those in \cite{Sorrentino}.

\section{Conclusions}

We have studied the problem of network synchronizability, which is
directly related to many other important problems as efficient
communication, searchability in the presence of congestion, many
computer-science tasks, etc. While generically, the optimal networks
for synchronizability, assuming un-weighted and un-directed links, are
super-homogeneous, {\it entangled} topologies, in which all nodes look
very much alike, in this work we have investigated the nature of
optimal scale-free networks. The final goal is to analyze how are the
degree-degree correlations of optimal scale-free networks.

For that, we have used the standard spectral approach consisting in
relating the degree of synchronizability to the Laplacian matrix
eigenratio $Q$. In a first part of this work, we have derived a
rigorous lower bound for $Q$ in terms of the correlation coefficient
$r$ (as defined by Eq.(\ref{r})), which is a measure of the
degree-degree correlations.  This lower bound turns out to be
proportional $1/(1-r)$, hence, showing that the more negative $r$
(i.e.  the more disassortative the network) the smaller the lower
bound and, therefore, the smaller values $Q$ is allowed to take, and
the better the synchronizability of the resulting network.

In the second part, we have explicitly constructed optimal networks
(with a fixed number of nodes, links, and a given scale-free degree
distribution) by employing a recently introduced simulated-annealing
algorithm \cite{Entangled,Ramanujan}. We find that optimal networks
tend to be disassortative, as found already in previous studies
\cite{Sorrentino}, and in agreement with the expectations from the
previously found lower bound.  However, as there is not a one-to-one
correspondence between $Q$ and the correlation coefficient $r$, more
disassortative networks do not always synchronize better.  Actually,
we have illustrated how during the optimization process, at some
point, the degree of assortativity increases (i.e. $r$ increases) as
the network becomes more and more synchronizable.  The emerging
optimal networks exhibit also a tendency to have large loops and a
rather intricate structure (as occurs for entangled networks).

Finally, we have reverted the optimization process and, by minimizing
$-Q$, we have found what we call pessimally synchronizable networks.
These topologies are characterized by a long string ended by two
``heads'' of nodes with degrees larger than $2$ (required to preserve
the scale-free degree distribution) and are, therefore, highly
assortative. Contrarily to the case above, loops are very short.
These topologies are the worst possible ones (compatible with the
imposed scale-free degree distribution) if the task is to synchronize
the network. But, on the contrary, they constitute the best choice if
the goal is to avoid synchronization (or, equivalently, avoid
communicability, searchability, homogenization), which might be
important for some applications.  For instance, in order to maximize
the average time that a random infection (or random walk) takes to
reach an arbitrary target node, this is the type of network to design.

\vspace{0.25cm} We acknowledge financial support from the Spanish
Ministerio de Educaci\'on y Ciencia (FIS2005-00791) and Junta de
Andaluc{\'\i}a (FQM-165).


\vspace{0.25cm}
\section*{References}
\begin{thebibliography}{10}

\bibitem{Reviews} S. H. Strogatz, Nature {\bf 410}, 268 (2001).\\
  A. L. Barab\'asi and R. Albert, Rev. Mod. Phys. {\bf 74}, 47 (2002). \\
  S. N. Dorogovtsev and J. F. F. Mendes, {\it Evolution of Networks:
    From
    Biological Nets to the Internet and WWW}, Oxford Univ. Press (2003).\\
  M. E. J. Newman, SIAM Review {\bf 45}, 167 (2003).\\ S. Boccaletti,
  V. Latora, Y. Moreno, M. Chavez, and D.-U. Hwang, Phys. Rep. {\bf
    424}, 175 (2006).


\bibitem{Pecora} M. Barahona and L. M. Pecora, Phys. Rev. Lett. {\bf
    89}, 054101 (2002).\\ See also, L. M. Pecora and T. L.  Carroll,
  Phys. Rev. Lett. {\bf 64}, 821 (1990); ibid, {\bf 80}, 2109 (1998).
  \\ X. F. Wang and G. Chen, Int. J. Bifurcation Chaos Appl.
  Sci. Eng. {\bf 12}, 187 (2002).
%X. F. Wang and G. Chen, IEEE Trans. Circuits and Systems I {\bf 49},
%54 (2002).
%In this paper the case where the stability
%interval is unbounded from above is studied. The larger stability
%range (maximum synchronizability) is obtained my maximizing the
%spectral gap.


\bibitem{Nish}
     T. Nishikawa, A. E. Motter, Y.-C. Lai, and F. C. Hoppensteadt,
%     Heterogeneity in oscillator networks: are smaller worlds easier to
%     synchronize?
     Phys. Rev. Lett. {\bf 91}, 014101 (2003).

\bibitem{Hong} H. Hong, B. J. Kim, M. Y. Choi, and H. Park,
%     Factors that predict better synchronizability on complex networks
      Phys. Rev. E. {\bf 69}, 067105 (2004). H. Hong, M. Y. Choi, and
      B. J. Kim, Phys. Rev. E {\bf 65}, 026139 (2002); Phys. Rev. E
      {\bf 65}, 047104 (2002).



\bibitem{Motter} A. E. Motter, C. Zhou, and J. Kurths, Phys. Rev E {\bf 71},
016116 (2005); AIP Conference Proceedings {\bf 776}, 201 (2005);
Europhys. Lett. {\bf 69}, 334 (2005). \\ T. Nishikawa and A. E. Motter,
Phys. Rev. E {\bf 73}, 065106 (2006).\\ A. E. Motter, New J. Phys. {\bf 9},
182 (2007).

\bibitem{Bocca} D.-U. Hwang, M. Chavez, A. Amann, and S. Boccaletti,
Phys. Rev. Lett. {\bf 94}, 138701 (2005). \\ M. Chavez, D.-U. Hwang, A. Amann,
H. G. E. Hentschel, and S. Boccaletti, Phys. Rev. Lett.  {\bf 94}, 218701
(2005).\\ M. Chavez, D.-U. Hwang, and S. Boccaletti, Eur.
Phys. J. Special Topics {\bf 146}, 129 (2007).


%\bibitem{Korniss} G. Korniss, Phys. Rev E {\bf 75}, 051121 (2007).

\bibitem{Atay} F. M. Atay, T. Biyikoglu, and J. Jost,
%On the synchronization of Networks with prescribed degree distributions
%arXiv:nlin.AO/0407024
Physica D {\bf 224}, 35 (2006).




\bibitem{Entangled} L. Donetti, P. I. Hurtado and  M. A. Mu\~noz,
%{\it Entangled networks, super-homogeneity, and the optimal network
%topology}, 
Phy. Rev. Lett.  {\bf 95}, 188701 (2005).\\
L. Donetti, P. I. Hurtado and M. A. Mu\~noz, 
%{\it Synchronization in
%Network Structures: Entangled Topology as Optimal Architecture for
%Network Design}, in ICCS 2006, Part III, LNCS 3993,
%Ed. V. N. Alexander et al.  
Lect. Notes in Comp. Sci.  {\bf 3993}, 1075 (2006).

\bibitem{Ramanujan}
L. Donetti, F. Neri, and M. A. Mu\~noz, 
%{\it Topological features of
%optimal networks: Expanders, Ramanujan graphs, Cage-graphs, entangled
%networks and all that}, 
J. Stat. Mech.: Theor. Exp. (2006) P08007.


\bibitem{Catalans} R. Guimera, A. Arenas, A. Diaz-Guilera, F. Vega-Redondo,
     and A. Cabrales,
%     Optimal network topologies for local search with congestion
      Phys. Rev. Lett. {\bf 89}, 248701 (2002).
%See also, I. Vragovi\'c, E. Louis, and A. Diaz-Guilera,
%Cond-mat/0410174.
%D. J. Ashton, T. C. Jarrett, and N. F. Johnson,
%Phys. Rev. Lett. {\bf 94}, 058701 (2005).


\bibitem{Maritan}
V. Colizza, J. R. Banavar, A. Maritan, and A. Rinaldo,
Phys. Rev. Lett. {\bf 92}, 198701 (2004).


\bibitem{Neural} J. J. Torres et al.
% M. A. Mu\~noz, J. Marro, and P. L. Garrido,
%{\it Influence of Topology on a Neural Network performance}.
Neurocomputing {\bf 58-60}, 229 (2004). \\
%; and
%references therein.
% D. Stauffer, et al.
% A. Aharony, L. da Fontoura Costa, and J. Adler,
%Eur. Phys. J. B {\bf 32}, 395 (2003).
%B. J. Kim, 
% Performance of networks of artificial neurons: the role of clustering
%Phys. Rev. E {\bf 69}, 045101(R) (2004). \\
G. Grinstein and R. Linsker, 
Proc. Natl. Acad. Sci. USA, {\bf 102}, 9948 (2005).


\bibitem{dynamics} 
%Paths to Synchronization on Complex Networks
J. G\'omez-Garde\~nes, Y. Moreno, and A. Arenas,
Phys. Rev. Lett.  {\bf 98}, 034101 (2007). \\
%A. Arenas and A. D{\'\i}az-Guilera,
%Synchronization and modularity in complex networks 
%Eur. Phys. J. Special Topics, {\bf  143}, 19  (2007).
J. G\'omez-Garde\~nes, Y Moreno, and A Arenas,
%Synchronizability determined by coupling strengths and topology on complex networks 
Phys. Rev. E {\bf 75}, 066106 (2007). \\
P. N. Graw and M. Menzinger, Phys. Rev. {\bf 72}, 015101 (2005).\\
E. Oh, K. Rho, H. Hong, and B. Khang,  Phys. Rev. E {\bf 72}, 047101 (2005).
%C. Zhou, and J. Kurths, Chaos, {\bf 16}, 015104 (2006).\\
%J. Ren, H. Yang, and Y.-C. Zhang, arXiv:cond-mat/0703232.

\bibitem{eigen} Note that the Laplacian eigenvalues $\lambda_i$
  satisfy $  0 = \lam_1 \le \lam_2 \le \ldots \le \lam_N \le 2k_{max}$ ,
where $k_{max}$ is the largest degree in the graph \cite{Graphs,Mohar}.


\bibitem{Mohar} B. Mohar, in {\it Graph Theory, Combinatorics, and
Applications, Vol 2}, Ed. Y. Alavi, G.  Chartrand, O. R. Oellermann, and A. J
Schwenk, Wiley, New York, 1991. pp. 871.

\bibitem{Graphs} B.  Bollob\'as, {\it Extremal Graph Theory} Academic
  Press, New York. 1978. \\
W. Tutte, {\it Graph Theory As I Have Known
    It}, Oxford U. Press, New York, (1998). \\ F. Chung, {\it Spectral
    Graph Theory}, Number 92 in CBMS Region Conference Series in
  Mathematics. Am. Math.  Soc. 1997.  


\bibitem{SA} %S. Kirkpatrick, C. D. Gelatt, and M. P. Vecchi, 
%Science {\bf 220} 671 (1983). \\ 
N. Metropolis, A. W. Rosenbluth,
M. N. Rosenbluth, A. H. Teller, and E. Teller, J. Chem. Phys. {\bf 21}
1087 (1953).\\
T. J. P. Penna, Phys. Rev. E {\bf 51}, R1 (1995).


\bibitem{Sarnak} 
%P. Sarnak,
% What is an expander?
%Notices Amer. Math. Soc. {\bf 51}, 762 (2004). \\ 
G. Davidoff, P. Sarnak and
A. Valette, {\it Elementary Number Theory, Group Theory and Ramanujan
Graphs}, London Math. Soc. Students Texts, 55, Cambridge (2003).

\bibitem{Estrada} E. Estrada, Phys. Rev. E {\bf 75}, 016103 (2007); 
J. Theor. Biol. {\bf 244}, 296 (2007). 



\bibitem{Sorrentino} M. di Bernardo, F. Garofalo, and F. Sorrentino,
  Proceedings of the 44th IEEE
  Conference on Decision and Control, pp. 4616 (2005). \\
%M. di Bernardo. F. Garofalo, and F. Sorrentino, arXiv:cond-mat/050623. 
F. Sorrentino, M. di Bernardo, G. Huerta, C\'uellar, and S. Boccaletti,
Physica D {\bf 224}, 123 (2006).  \\
See also, B. Wang, H. Tang, T. Zhou, and
Z. Xiu, arXiv:cond-mat/0512079.

\bibitem{asdis} M.E.J. Newman, Phys. Rev. Lett. {\bf 89}, 208701 (2002).

\bibitem{Glass} L. Glass, Nature {\bf 410}, 277 (2001).


\bibitem{GN} M. Girvan, M. E. J. Newman,
Proc. Natl. Acad. Sci. USA, {\bf 99}, 7821-7826 (2002).
M. E. J. Newman, M. Girvan, Phys. Rev. E {\bf 69}, 026113 (2004).
See also, L. Donetti and M. A. Mu\~noz,
%{\it Detecting Network Communities: a
%new systematic and powerful algorithm},
J. Stat. Mech.: Theor. Exp. (2004) P10012;
L. Donetti and M. A. Mu\~noz, 
%{\it Improved spectral algorithm for the detection of network Communities},
in ``{\it Modeling Cooperative behavior in the social sciences}'', AIP
Conf. Proc. 779, 104 (2005).


\bibitem{conf} See, M. Molloy and B. Reed, Comb. Prob. and Comput.
  {\bf 6}, 161 (1995).\\
  M. Molloy and B. Reed, Comb. Prob. and Comput. {\bf 7}, 295 (1998).


\end{thebibliography}
\end{document}